\documentclass[aps,prl,twocolumn,groupedaddress]{revtex4-1}

\usepackage[dvipdfmx]{graphicx}
\usepackage{bm}
\usepackage{siunitx}
\newcommand{\keff}{k_{\mathit{eff}}}

\DeclareSIUnit\gauss{G}
\newcommand\fig[1]{Fig.~\ref{fig:#1}}

\usepackage[dvipdfmx]{hyperref}
\hypersetup{ colorlinks=true,linkcolor=red,citecolor=blue,urlcolor=blue}
\begin{document}


\title{Atom interferometry in an optical cavity}


\author{Paul Hamilton}
\email[]{paul.hamilton@berkeley.edu}
\author{Matt Jaffe}
\author{Justin M. Brown}
\altaffiliation{Present address: C.S. Draper Laboratory, Inc., 555 Technology Square, Cambridge, Massachusetts 02139, USA}
\author{Lothar Maisenbacher}
\altaffiliation{Present address: Max-Planck-Institut f\"ur Quantenoptik, 85748 Garching, Germany}
\author{Brian Estey}
\author{Holger M\"uller}
\altaffiliation{Also at Lawrence Berkeley National Laboratory, One Cyclotron Road, Berkeley, California 94720, USA.}
\affiliation{Department of Physics, University of California, Berkeley, California 94720, USA}

\date{\today}

\begin{abstract}
We propose and demonstrate a new scheme for atom interferometry, using light pulses inside an optical cavity as matter wave beamsplitters.  The cavity provides power enhancement, spatial filtering, and a precise beam geometry, enabling new techniques such as low power beamsplitters ($<\SI{100}{\micro\watt}$), large momentum transfer beamsplitters with modest power, or new self-aligned interferometer geometries utilizing the transverse modes of the optical cavity.  As a first demonstration, we obtain Ramsey-Raman fringes with $>75\%$ contrast and measure the acceleration due to gravity, $g$, to $\SI{60}{\micro g\per\sqrt{\hertz}}$ resolution in a Mach-Zehnder geometry.  We use $>10^7$ cesium atoms in the compact mode volume ($\SI{600}{\micro\meter}$ $1/e^2$ waist) of the cavity and show trapping of atoms in higher transverse modes. This work paves the way toward compact, high sensitivity, multi-axis interferometry.
\end{abstract}


\maketitle

\begin{figure}%
\includegraphics[width=\columnwidth]{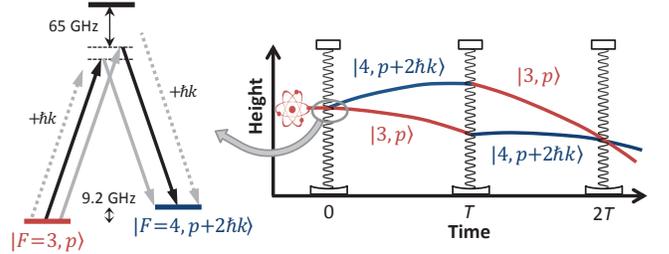}%
\caption{Left: Energy level scheme. A two-photon Raman interaction between the ground hyperfine states of cesium transfers momentum to the atoms. The laser (black arrows) is modulated at the hyperfine frequency creating sidebands (gray arrows).  Two resonant pathways (solid arrows) can interfere depending on the position of the atoms and the detuning from an optical cavity resonance.  Right:  Mach-Zehnder interferometer.  Momentum transfer from three pulses of a laser standing wave (wavy black lines) separated by a time $T$ split, redirect, and interfere a matter wave (red and blue lines).}
\label{fig:ramanmz}%
\end{figure}

In a light-pulse atom interferometer, recoils from photon-atom interactions are used to split and interfere matter waves (see \fig{ramanmz}). These interferometers have been used to measure the gravitational acceleration $\vec g$ \cite{Peters2001}, rotation $\vec\Omega$ \cite{Durfee2006}, gravity gradients \cite{McGuirk2002}, the fine structure constant \cite{Bouchendira2011}, Newton's gravitational constant \cite{Fixler,TinoG}, and absolute masses in a proposed revision of the SI \cite{CCC,Biraben2013}; to test Einstein's equivalence principle \cite{Fray2004,Bonnin2013,TinoEP,Schlippert2014}; and have been proposed to measure the free fall of antimatter \cite{AntiH} and to detect gravitational waves \cite{KasevichWaves,Mango,Barrett2013}. The sensitivity of a conventional Mach-Zehnder interferometer increases with the measured phase difference
\begin{equation}
\phi = (2\vec{\Omega}\cdot[\vec\keff \times(\vec{v}_0 +\vec{g}\,T)] +\vec \keff \cdot\vec{g})T^2
\label{eq:MZphi}
\end{equation}
(where $\vec{v}_0$ is the initial velocity of the atom), which scales with the pulse separation time $T$ and the recoil momentum $\vec{p} = \hbar\vec\keff$, where $\vec\keff$ is the effective wavenumber of the photons. State of the art atom interferometers are limited by several engineering boundaries. $T$ is limited by the free-fall time in atomic fountains, which are now as high as \SI{10}{\meter} \cite{Sugarbaker2013,Zhou2011}.  Multiphoton interactions can increase the recoil momentum to a multiple $n\hbar k$ of the single photon recoil \cite{BraggPRL,Chiow2011,BBB,CladeBloch,McDonald2013} but are limited by the available laser power (e.g. \SI{6}{\watt} in \cite{6Wlaser}, \SI{43}{\watt} in \cite{Chiow2012}).  Finally, wavefront distortions spread the local wavevector around its mean, lowering interference contrast and reducing both sensitivity and accuracy.  An optical cavity can solve these problems by providing spatial filtering to clean the wavefronts and enhancing laser intensity.   However, running an atom interferometer inside an optical cavity presents challenges in keeping the atoms in the relatively small cavity mode volume and having multiple laser frequencies (needed due to recoil frequency shifts, Doppler shifts, and atomic structure) simultaneously resonant with the cavity. Here, we present a cesium atom interferometer inside an in-vacuum optical cavity and demonstrate gravity measurements using less than $100\mu$W of laser power incident on the cavity.  

The use of an optical cavity has many advantages.  First, laser power limits interferometers using both large momentum transfer beamsplitters and optical lattices.  For example, Bragg diffraction requires an intensity proportional to $n^2$ for constant pulse duration, or $n^4$ for constant single-photon scattering rate at an increased detuning \cite{Losses}. With resonant enhancement in a cavity, we may achieve $n=50-100$-photon Bragg transitions using tens of milliwatts of power from a standard diode laser as opposed to the multiple watt systems recently developed \cite{6Wlaser,Chiow2012}.   Similarly, we can reduce scattering from optical lattices by using increased intensity at a larger detuning.

Second, spatial variations in the laser amplitude and phase can reduce interferometric contrast by causing imperfect beamsplitters or unwanted forces in schemes which use an optical lattice to hold atoms  \cite{Clade2010,Bresson2012}. Using the optical cavity as a spatial filter gives well-defined optical phase fronts, increasing the interferometer contrast and reducing systematic errors. 

Third, interferometer geometries with enclosed spatial area typically use independent laser beams, whose relative vibration and alignment need to be tightly controlled. Using multiple transverse spatial modes of the optical cavity provides self-aligned interferometry beams.  In addition, multiple simultaneous interferometers could provide common-mode rejection of vibrational noise in rotation measurements. 
 
Fourth, many systematics, e.g. gravity gradients, electric and magnetic fields and gradients, are more easily controlled in a small volume. The combination of large momentum transfer and long coherence times in an optical lattice provides high sensitivity in a compact area. 

Finally, the position uncertainty typical in atomic fountains, optical wavefront curvature, and Gouy phase shifts are leading systematics in precision experiments \cite{Fixler,Bouchendira2011,CCC,TinoG}.  The well-defined geometry of the optical cavity reduces them.  Counterpropagation of the interferometry beams is also intrinsic in an optical cavity.  Optical beam parameters can be determined precisely by measuring the transverse mode spacing of the cavity.

Our experiment uses a two-dimensional magneto-optical trap (2D-MOT) to feed a 3D-MOT through a differential pumping stage. A vacuum of below $10^{-9}$ Torr is maintained inside the main vacuum chamber. All frequencies are stabilized (``locked") to a reference laser, which is in turn stabilized to a cesium transition by modulation transfer spectroscopy.  An experimental run starts with loading $\approx\num{5e8}$ cesium atoms into the 3D-MOT in one second. After increasing the detuning of the 3D-MOT beams from $-1.5$ to $-12$ linewidths ($\Gamma=2\pi\times\SI{5.2}{\mega\hertz}$) and decreasing the optical power, polarization gradients cool the atoms in the $F = 4$ state to \SI{6}{\micro\kelvin}.

\begin{figure}%
\includegraphics[width=\columnwidth]{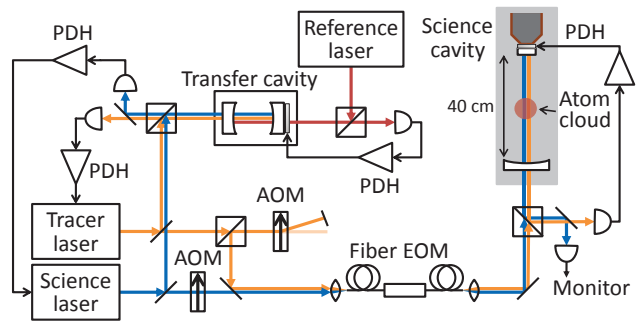}%
\caption{Cavity and laser frequency stabilization.  An external transfer cavity acts as a common reference for both the science laser and a second far-detuned laser (the ``tracer" laser) which is used to stabilize the science cavity inside the vacuum chamber (gray box).  The transfer cavity itself is stabilized to a reference laser.  Cavity lengths and laser frequencies are stabilized via feedback using the Pound-Drever-Hall (PDH) method.}
\label{fig:locks}%
\end{figure}

For operation of the interferometer, both frequencies needed to drive Raman or Bragg transitions (the ``science laser") must be kept on resonance with the ``science" cavity surrounding our interferometer.  We use a second ``tracer" laser to stabilize the length of the science cavity.  The tracer wavelength of \SI{780}{nm} is far from any transition in cesium and has a negligible effect on the atoms.   Both lasers are locked to an external transfer cavity whose length is stabilized to the reference laser (see \fig{locks}).  Before going to the science cavity, the tracer laser is double-passed through a 200 MHz bandwidth acousto-optic modulator (AOM, Brimrose TEF-300-200-0.780) which is tuned such that both lasers are simultaneously resonant with both cavities. Feedback to the science cavity is applied to a piezo-driven $1/2''$ diameter flat gold mirror with a feedback bandwidth of \SI{40}{\kilo\hertz} \cite{YePiezo}.  The other cavity mirror has a $1''$ diameter and \SI{5}{\meter} radius of curvature.  The fundamental longitudinal mode of the cavity has a beam waist of \SI{600}{\micro\meter} located at the surface of the flat mirror, a finesse of $\approx150$, and a linewidth of $\SI{2.5}{\mega\hertz}$.  The transverse modes of the cavity are non-degenerate in resonance frequency.  For the demonstrations below, where the laser is tuned to only couple strongly to the fundamental mode of the cavity, the incoming interferometry beam is thus effectively spatially filtered.

The length of the cavity (\SI{40.756}{\centi\meter}) was chosen to be nearly an integer number of half wavelengths of the ground state hyperfine splitting $\nu_{\rm HFS}=\SI{9192631770}{\hertz}$ of cesium.  This allows the two science-laser frequencies needed for Raman transitions to be simultaneously near-resonant with the cavity mode. They are generated from a single laser by a fiber coupled broadband EOM (Eospace). Low phase noise of this frequency is crucial, as it directly enters the measured phase of the interferometer. We use a dielectric resonator oscillator locked to a harmonic of a low-noise quartz crystal. Low phase noise is achieved by using a nonlinear transmission line (Picosecond Pulse Labs) as a harmonic generator. For Bragg diffraction and Bloch oscillations, the science laser passes through an AOM which can be modulated at two frequencies \cite{SCI}.

\begin{figure}%
\includegraphics[width=225pt]{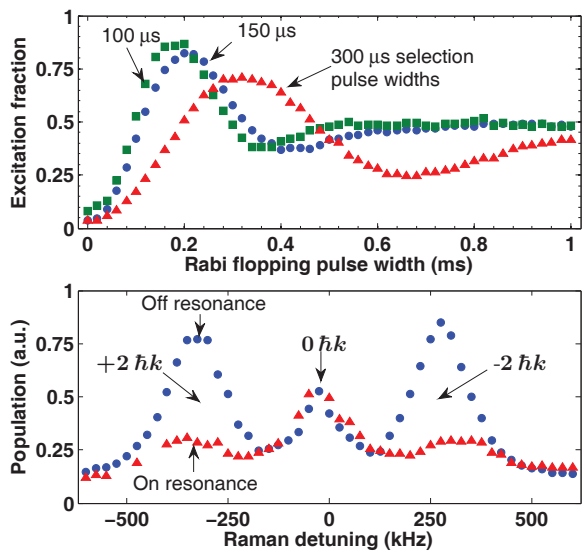}
\caption{
Top: Rabi flopping after three state selection pulses with fixed intensity having pulse widths of \SI{300}{\micro\second} (red), \SI{150}{\micro\second} (blue), and \SI{100}{\micro\second} (green), followed by clearing pules shows improved contrast as atoms near the center of the cavity mode are preferentially selected. Bottom: Cavity suppression of velocity sensitive transitions. Changing the cavity length to be either on resonance with the carrier (red) or \SI{2.2}{\mega\hertz} off resonance (blue) emphasizes one carrier-sideband pair and prevents cancellation of the transition amplitude.}%
\label{fig:stateselect}%
\end{figure}

We coarsely select atoms from the MOT that are located in the science cavity mode by first turning on the science laser, detuned $\approx\SI{{-}65}{\giga\hertz}$ from the $\emph{D}_2$ transition, to create an optical lattice.  Atoms not located in the lattice or with too high temperature fall away, leaving $\num{2e8}$ atoms.  The temperature transverse to the cavity mode increases to $\SI{25}{\micro\kelvin}$ as the atoms are loaded into the lattice, but the longitudinal temperature is reduced to $<\SI{2}{\micro\kelvin}$ upon release from the lattice, most likely by adiabatic expansion. We then optically pump $>75\%$ of the atoms to the magnetically insensitive $F=4,m_F=0$ sublevel by applying a magnetic bias field of $\SI{140}{\milli\gauss}$ and a retroreflected beam resonant with the $F=4\rightarrow 4$ transition with linear polarization parallel to the bias field. A small amount of repump light on the $F=3 \rightarrow 4$ transition is simultaneously applied. For both state and spatial selection, we apply a series of Raman pulses on the $F=3,m_F =0\leftrightarrow F=4,m_F=0$ clock transition.  Each pulse is followed by a clearing beam (either on the cycling $F=4\rightarrow 5$ or $F=3\rightarrow 2$ transitions). This both removes atoms that were not initially in the $m_F=0$ sublevel and preferentially selects atoms towards the center of the cavity mode.  Without this spatial selection the widths of the atomic distribution and optical interferometry beams are initially similar and we can achieve a maximum population transfer of only $\approx 50\%$ with low contrast on Rabi oscillations.  However, using three state selection pulses with a length that maximizes population inversion at the center of the cavity mode allows us to reach $90\%$ population transfer (see \fig{stateselect}, top).

Use of an EOM to generate the Raman frequency pair is a simple way to reach low phase noise, but leads to two sidebands of equal amplitude and opposite phase.  At detunings large compared to the hyperfine structure of the excited state the two sideband/carrier pairs drive Raman transitions (see \fig{ramanmz}, left) which interfere with a position dependent phase. Because the hyperfine splitting is $\approx\SI{2.2}{\mega\hertz}$ off (similar to the science cavity line width) from being a multiple of the free spectral range of the science cavity, we may selectively enhance one carrier/sideband pair while suppressing the other, preventing this cancellation.  

\begin{figure}%
\includegraphics[width=225pt]{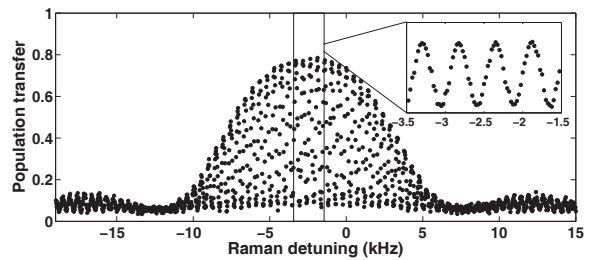}%
\caption{Raman-Ramsey fringes.  Two Raman $\pi/2$ pulses on the cesium clock transition are applied with a separation time $T=\SI{2}{\milli\second}.$  Each point is the average from four experimental runs.  The high contrast and signal-to-noise of the fringes demonstrates that, despite the small size of the optical cavity mode waist, spatial selection of the atoms gives a uniform Rabi frequency for the $\num{2e7}$ atoms participating in the interferometer sequence.}%
\label{fig:ramsey}%
\end{figure}

We now present experiments that lead up to our demonstration of intra-cavity atom interferometry: velocity-insensitive and -sensitive Raman pulses, a Raman-Ramsey sequence and, finally, atom interferometry measuring the acceleration due to free fall, $g$.  Raman transitions using a copropagating pair of photons are velocity insensitive and transfer negligible momentum to the atoms.  We optically drive a combination of two such Raman $\pi/2$ pulses between the clock states $F=3, m_F=0 \rightarrow F=4, m_F=0$, spaced by a pulse separation time of T = 2\,ms to generate the Ramsey fringes shown in \fig{ramsey}.  The high contrast demonstrates the good coherence of the process despite the comparable size of the atomic cloud and interferometry beams.  This measurement is an optical version of a cesium fountain clock.  The optical pulses lead to an AC Stark shift of the transition frequency (e.g. \SI{-2.35}{\kilo\hertz} in \fig{ramsey}).  By varying the optical intensity while keeping the pulse area constant we find an extrapolated value for the transition frequency at zero intensity of \SI{9192631590\pm50}{\hertz}.  With microwave pulses we find that, after a \SI{10}{\hertz} correction due to the quadratic Zeeman effect, the measured transition frequency of \SI{9192631770\pm1}{\hertz} agrees with the international definition of the second.

Next, we demonstrate velocity-sensitive Raman pulses (using counterpropagating pairs of photons) that transfer a momentum of $\pm2\hbar k$ (\fig{stateselect}, bottom).  The falling atoms cause a Doppler shift of the velocity-sensitive transitions which allows us to suppress the unwanted velocity-insensitive transition.  As discussed above, detuning the cavity resonance to suppress one EOM sideband further enhances the velocity-sensitive transition.

\begin{figure}%
\includegraphics[width=225pt]{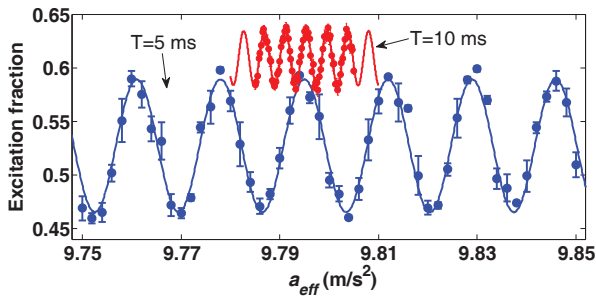}%
\caption{Mach-Zehnder fringes of a $\pi/2-\pi-\pi/2$ pulse sequence on the cesium clock transition with pulse separation times $T=\SI{5}{\milli\second}$ (blue) and $T=\SI{10}{\milli\second}$ (red).  The Raman frequency difference is linearly ramped to give an effective acceleration, $a_{\mathit{eff}}$, which compensates for the Doppler shift from the acceleration due to gravity.  The phase of the interferometer is given by $\keff(g-a_{\mathit{eff}})T^2$.}%
\label{fig:gfringes}%
\end{figure}

Finally, we demonstrate atom interferometry and perform a gravity measurement by a $\pi/2-\pi-\pi/2$ combination of three velocity sensitive Raman pulses, which constitutes a Mach-Zehnder interferometer.  To remain resonant as the freely falling atoms accelerate, the difference frequency in the Raman frequency pair is swept at a rate of  $\keff a_{\mathit{eff}}$ ($\approx2\pi\times\SI{23}{\mega\hertz\per\second}$ for $a=\SI{9.8}{\meter\per\second\squared}$). We detect the interferometer outputs separately by first pushing atoms in either $F=4$ or $F=3$ to the side with our clearing beams and then using fluorescence detection on a CCD camera to spatially resolve the two populations. After normalization to take out atom-number fluctuations, we obtain the interference fringes shown in \fig{gfringes} by scanning the rate of the frequency ramp.  When the frequency ramp matches the acceleration the interferometer phase, $\keff(g-a_{\mathit{eff}})T^2$, should be zero independent of the pulse separation time $T$.  At a maximum pulse separation time $T=\SI{15}{\milli\second}$ we achieve a resolution of $\SI{60}{\micro g\per\sqrt{\hertz}}$, similar to the sensitivity achieved by other compact atom interferometers \cite{McGuinness2012}.  Additionally, we find that, with as little as \SI{87}{\micro\watt} of power, and at a smaller single photon detuning of $\SI{2}{\giga\hertz}$, we attain fringes with $>10\%$ contrast for $T=\SI{1}{\milli\second}$. 

We have demonstrated the first atom interferometer in an optical cavity. We have shown that tuning the cavity resonance allows us to selectively address both velocity-sensitive and -insensitive transitions as well as to choose the direction of the momentum kicks in the interferometer, a technique that can be used to reverse many systematic effects. Atoms have been loaded efficiently into the fundamental cavity mode. The setup uses diode lasers only. Using sidebands generated by a wideband EOM to drive Raman transitions avoids the noise usually contributed by a laser phase lock, and is possible without undue systematic effects thanks to the cavity's rejection of the unwanted sideband.  Despite the small size of the cavity mode we achieved high contrast using optical beamsplitters.

\begin{figure}%
\includegraphics[scale=0.55]{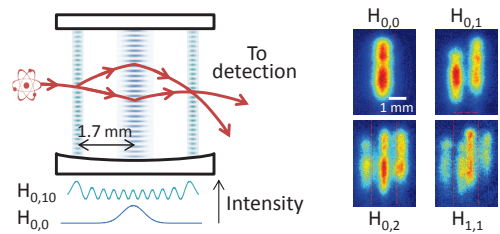}%
\caption{Left:  A possible implementation of a rotation sensitive interferometer which encloses a spatial area using the transverse Hermite-Gaussian $H_{0, 0}$ and $H_{0, 10}$ modes of the cavity.  Right: Fluorescence images of atoms in optical lattices formed by transverse modes of the optical cavity.}%
\label{fig:transverse}%
\end{figure}

While this proof-of-principle demonstration relied on hyperfine changing Raman transitions, we expect with the power enhancement of the optical cavity that large momentum transfer multiphoton Bragg beamsplitters will be possible \cite{Losses}. For the future, the combination of very large momentum transfer and short pulse separation times will make a compact setup which is less sensitive to low-frequency vibrations. Localization of atoms in the cavity, and precise control of the mode of the laser beam should help suppressing major systematics in measurements of the fine structure constant, a proposed mass standard \cite{CCC}, the gravitational constant \cite{Fixler,TinoG}, or deviations from Newtonian gravity at short distances \cite{Wolf2007}. With these beamsplitters and intracavity optical lattices we plan on exploring a range of applications from inertial sensors \cite{Hamilton2014} to the first measurement of the gravitational analog of the Aharonov-Bohm effect \cite{Hohensee2012}.  For example, in \fig{transverse} we show one possible realization of a rotational sensor utilizing self-aligned transverse modes of the cavity as well as a demonstration of atoms loaded into such modes.  Recently developed methods for both preparing and detecting spin squeezed states \cite{Bohnet2014} in a similar optical cavity could lead to improved sensitivity beyond the standard quantum limit.  In addition, the compact size and low power requirements make a cavity interferometer ideal for future mobile or space-based applications \cite{Aguilera2014}.  Finally, the techniques discussed here may be of interest to proposed experiments using optical cavities to enhance diffraction in systems without easily accessible resonant transitions, e.g. macroscopic masses \cite{Asenbaum2013} or antihydrogen \cite{AntiH}. 

\begin{acknowledgments}
We thank E. Simmons and T. Gavrilchenko for their contributions to the apparatus and P. Haslinger, D. Schlippert, and C. Thomas for their comments on the manuscript.  J.M.B. thanks the Miller Institute for Basic Research in Science for support. This work has been supported by the David and Lucile Packard Foundation, the Defense Advanced Research Projects Agency, the National Aeronautics and Space Agency and the National Science Foundation.
\end{acknowledgments}

\bibliography{Bibliography/cavityai}

\end{document}